\documentclass[amsmath,amssymb,superscriptaddress,showpacs,twocolumn]{revtex4-2}
\usepackage{graphicx}
\usepackage{dcolumn}
\usepackage{bm}
\usepackage[colorlinks=true, citecolor=blue, urlcolor = blue, linkcolor= red, bookmarks=true]{hyperref}

\begin{document}

\title{Rotating black holes in general relativity coupled to nonlinear electrodynamics}
\author{Sushant~G.~Ghosh } \email{sghosh2@jmi.ac.in}
\author{Rahul Kumar Walia } \email{rahul.phy3@gmail.com}
\affiliation{Centre for Theoretical Physics, Jamia Millia Islamia, New Delhi 110 025, India}
\affiliation{Astrophysics Research Centre, School of Mathematics, Statistics and Computer Science, University of KwaZulu-Natal, Private Bag 54001, Durban 4000, South Africa}
\date{\today}

\begin{abstract}
We find an exact spherically symmetric magnetically charged black hole solution to general relativity (GR) coupled to nonlinear electrodynamics (NED) with an appropriate Lagrangian density. In turn, starting with this spherical black hole as a seed metric, we construct a rotating spacetime, a modification of Kerr black hole, using the revised Newman-Janis algorithm that depends on mass, spin, and a NED parameter $g$. We find an exact expression for thermodynamic quantities of the black holes like the mass, Hawking temperature,  entropy, heat capacity, and free energy expressed in terms of horizon radius, and they show significant deviations from the Kerr case owing to NED. We also calculate analytical expressions for effective Komar mass and angular momentum for the rotating black hole and demonstrate that the Komar conserved quantity corresponding to the null Killing vector at horizon obeys  $\mathcal{K}_{\chi}=2S_+T_+.$  The radiating counterpart renders a generalization of Carmeli's spacetime as well as Vaidya's spacetime in the appropriate limits.
\end{abstract}


\maketitle
\section{Introduction}
Black holes are fascinating theoretical predictions of Einstein's general relativity (GR) \cite{he}. The first exact solution to this theory was found by Schwarzschild \cite{Schwarzschild:1916uq}, describing a static spherically symmetric spacetime. The  astrophysically more relevant rotating black holes metrics are Kerr \cite{Kerr:1963ud}\ and Kerr-Newman \cite{nja}, which are undoubtedly the most effective exact solutions in the GR, and can arise as the final fate of gravitational collapse.  As predicted by GR, black holes have direct observational evidence of their existence after the Event Horizon Telescope (EHT) collaboration released the first-ever image of the supermassive black hole M87*. The spacetime singularities in GR are inevitable, which are also predicted by the celebrated singularity theorem \cite{Akiyama:2019cqa,Akiyama:2019bqs}.  

Whereas theoretically, black hole solutions are a suitable playground to test theories that modify GR,  where nonlinearities are involved. The behavior of charged particles around them may be described by nonlinear electrodynamics (NED); hence it is pertinent to consider a black hole in GR coupled to a NED \cite{Lammerzahl:2018zvb}.  This issue was with the invention of the Einstein-Born-Infeld (EBI) spacetime \cite{ebi}, which is one of the most used GR coupled NED models; it turns out that the specific values of its parameters can remove the curvature singularity present in the Reissner– Nordstrom black hole.  The interest in NED is also because, in the effective theory action arising from superstrings, a generalized BI action occurs naturally as the leading part \cite{string}. Indeed, the low energy effective action in an open superstring in loop calculations also lead to BI type actions \cite{string}. The effects of NED, from an astrophysical point of view, become quite crucial in super-strongly magnetized compact objects, such as pulsars and magnetars  \cite{magnetar}.  
Interestingly, Bardeen spacetime \cite{Bardeen}, the first regular black hole model, was also interpreted as a spherically symmetric solution to GR coupled to a NED \cite{ABG99}.  Subsequently, there has been intense activities in the investigation of NED sourced black holes \cite{sa,Hayward,Xiang,hc,lbev,Balart:2014cga}. However, the black hole spin plays a critical and key role in the astrophysical observations of any relativistic process,  which evoked generalization of the spherically symmetric EBI black hole \cite{Hoffmann} in the rotating case, Kerr-Newman-like solution, was studied by Lombardo \cite{CiriloLombardo:2004qw}, also of the rotating regular black holes \cite{Johannsen:2011dh,Bambi:2013ufa,Neves:2014aba,Toshmatov:2014nya, Ghosh:2014pba, Ghosh:2015ovj,Ghosh:2014hea}. These rotating black holes were obtained via the Newman-Janis algorithm \cite{Newman:1965tw} or other similar techniques \cite{Azreg-Ainou:2014aqa,Azreg-Ainou:2014nra,Azreg-Ainou:2014pra}. Because of their importance for astrophysical observations, these NED sourced black holes have been studied in varities of context \cite{Javed:2019ynm,Javed:2019rrg,Jusufi:2018kmk,Ovgun:2021ttv,Javed:2020lsg,Javed:2019kon,Ovgun:2021ttv,Kuang:2018goo,Kumar:2018ple}. Notably, the EHT shadow observations of the M87* black hole also do not altogether rule out these black holes \cite{M87}. Further, the no-hair theorem still lacks direct evidence \cite{no-hair}, and its actual nature has not yet known that opens the arena for investigating the properties for black holes that differ from Kerr black holes. The black holes, not a solution to Einstein's vacuum equations but of GR coupled to NED or modified gravity theory, can avoid the singularity theorems because they obey the weak energy condition, not the strong ones \cite{Bambi:2013ufa,Ghosh:2014pba,Toshmatov:2014nya}. 

Here, our aim is to find out new black hole configurations that may result from the GR coupled to the NED. Since it violates the energy conditions, the rotating metric will be of interest to black hole physics and cosmology. We first obtain a spherically symmetric black hole solution to GR coupled to NED, which is taken as a seed metric to construct a rotating spacetime using the revised Newman-Janis algorithm \cite{Azreg-Ainou:2014aqa,Azreg-Ainou:2014nra,Azreg-Ainou:2014pra}.  The considered non-rotating seed metric has several attractive features: the spacetime has the scalar polynomial singularity \cite{he} at $r=0$ and reflection symmetry, i.e., the transformation  $r \to - r$ leads to a replica of the same spacetime.  We note that the obtained rotating solution also retains this property and has significant deviations from the Kerr spacetime.  We demonstrate a black hole with two inner and outer horizons for a specific range of NED charge parameter, and thermodynamic quantities are also determined.

\section{Black hole solution}
\label{ }
We first obtain the spherically symmetric black hole metric that can be derived from the action of GR minimally coupled to the NED \cite{Balart:2014cga,Amir:2015pja}:
\begin{equation}\label{action}
	\mathcal{S}= \int \mathrm{d} ^4x \sqrt{-\mathbf{g}} \left(\frac{1}{16\pi}  R-\frac{1}{4\pi} \mathcal{L}(F)\right),
\end{equation}
where $R$ is the Ricci scalar, $\mathbf{g}$ is the determinant of the metric tensor,  and $\mathcal{L}(F)$ is the Lagrangian density of NED which is a function of $F=1/4 F^{\mu\nu}F_{\mu\nu}$ with $F_{\mu\nu}=\partial_{\mu}A_{\nu}-\partial_{\nu}A_{\mu}$, the field strength tensor of NED four-potential $A_{\mu}$. The Lagrangian $\mathcal{L}(F)$ is an arbitrary function of invariant $F$ which should have the Maxwell limit $\mathcal{L}(F) \to F$ in weak-field limit. On varying the action (\ref{action}), we obtain the following field equations 
\begin{eqnarray} \label{field_eq1}
	G_{\mu\nu} &=& T_{\mu\nu} \equiv  2 \left( \frac{\partial \mathcal{L}(F)}{\partial F} F_{\mu \lambda} F _{\nu}{}^{\lambda} - g_{\mu\nu} \mathcal{L}(F) \right), \label{field_eq2}
\end{eqnarray}
where $G_{\mu\nu}$ is the Einstein tensor  and the tensor $ F_{\mu\nu}  $ obeys the dynamic equation
\begin{eqnarray}\label{ee1}
	\nabla_{\mu}\left(\frac{\partial \mathcal{L}(F)}{\partial F}F^{\mu\nu}\right)=0,
\end{eqnarray}
and the Bianchi identities
\begin{eqnarray}\label{ee2}
	\nabla_{\mu}\left(^*F^{\mu\nu}\right)=0,
\end{eqnarray}
where $*$  denotes the Hodge dual. To find black hole solutions, it is convenient to choose a line element with a radial coordinate such that $g_{tt}=1/g_{rr}$. Accordingly, we choose the following ansatz:
\begin{equation}
	ds^2 = -g_{tt}(r)dt^2+g_{rr}(r) dr^2 + r^2 (d\theta^2+\sin^2\theta d\phi^2). 
	\label{metric}
\end{equation}
We consider the NED field tensor \cite{Ghosh:2020ece, Balart:2014cga,Neves:2014aba} 
\begin{equation}
	F_{\mu\nu}=2\delta^{\theta}_{[\mu}\delta^{\phi}_{\nu]}g(r)\sin\theta.
\end{equation} 
By using Eq. (\ref{ee1}), we get the condition $g'(r)\sin\theta d r\wedge d\theta \wedge d\phi=0$, which implies $g(r)=\text{constant}=g$, and we get
$ F_{\theta\phi}=g \sin\theta $\ {and}  $ F={g^2}/{2r^4} $. 
The solution that we are interested in can be obtained from the Lagrangian density
\begin{equation} \label{L_term}
	\mathcal{L}(F) = \frac{2 \sqrt{g}F^{5/4} }{s \left( \sqrt{2} + 2 g \sqrt{F}\right)^{3/2} },
\end{equation}
where $g$ is a magnetic charge and $s>0$ is a constant to be fixed later. Accordingly, the Eq.~(\ref{field_eq1}) for the NED-modified energy-momentum tensor, due to the spherical symmetry, leads to the following form $T_{\mu\nu}  = \mbox{diag} \left(- \rho, P_1, P_2, P_3 \right).$ Here, $P_1$ and $P_2$, respectively, are the radial and transverse pressures with $\rho = 2\mathcal{L}(F)$. Using the field  Eq.~(\ref{field_eq1}) with Lagrangian density given by Eq.~(\ref{L_term}) and $F$, after integration leads to the following metric coefficient
\begin{eqnarray}\label{solution}
g_{tt}(r) = g^{-1}_{rr}(r)=-\left(1 -\frac{2 M }{\sqrt{g^2+r^2}} \right),
\end{eqnarray}
Here, $M$ is the black hole mass and $s = g/2M$ is parameter related to NED. The metric (\ref{metric}) is solution of the  Einstein field equations (\ref{field_eq1}) with NED as source corresponding to the Lagrangian density (\ref{L_term}). The quantity $g$ parametrically controls deviations from  the Schwarzschild black hole solution and the limit $g \to 0$ provides a cross check for consistency, the GR results are recovered.  

We discuss the behavior of curvature invariants $ {{\tt `RicciSq `}}= R_{ab} R^{ab}$ ($R_{ab}$ is the
Ricci tensor) and the Kretschmann invariant $ {{\tt `K `}} = R_{abcd} R^{abcd}$ ($R_{abcd}$ is the Riemann tensor).  They are given by 
\begin{eqnarray*}
 && {{\tt `RicciSq `}} =  {\frac {8{M}^{2}{
 			g}^{4}}{ \left( {g}^{2}+{r}^{2} \right) ^{5}{r}^{4}} \left( {g}^{4}+2
 	\,{g}^{2}{r}^{2}+{\frac {13\,{r}^{4}}{4}} \right) }   ,\nonumber\\ 
&&  {{\tt `K `}} =  {\frac {16{M}^{2}}{ \left( {g}^{2}+{r}^{2}
		\right) ^{5}{r}^{4}} \left( {g}^{8}+4\,{g}^{6}{r}^{2}+{\frac {29\,{g}
			^{4}{r}^{4}}{4}}+5\,{g}^{2}{r}^{6}+3\,{r}^{8} \right) }
\end{eqnarray*}
For $M \neq0$, these invariants are well-behaved everywhere, except at $r=0$, where noteworthy they diverge.  Hence, the first surprising feature of the above spacetime geometry is that metric functions are regular, but  the spacetime has the scalar polynomial singularity \cite{he} at $r=0$. Another novelty of the solution is its reflection symmetry, the transformation  $r \to -r$ leads to a replica of the same spacetime. The exact solution (\ref{metric}) is similar to interesting black hole spacetime proposed as a toy model by Simpson and Visser \cite{Simpson:2019cer} wherein the angular part is different from the above solution, and their solution is regular, which interpolates between the standard Schwarzschild black hole and the Morris–Thorne traversable wormhole. 

The lack of rotating black hole models in modified gravities, which are typically helpful for astrophysical observation, substantially hinders testing modified gravities from observations.   It motivates us to seek a rotating generalization of the metric (\ref{metric}) or finding a Kerr-like metric, namely, a rotating black hole that may be helpful for astrophysical observations as spin plays an important role.
\begin{figure}
	\includegraphics[scale=0.77]{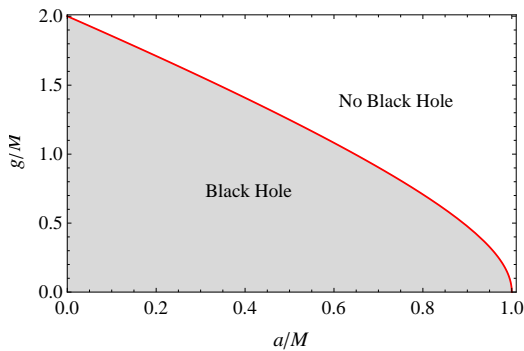}	
	\caption{The parameter plane for $(a,\;g)$  of the rotating   metric and  thin red line separates black holes from configurations without an event horizon.} \label{plot1}	
\end{figure}

\subsection{Rotating Metric}
The Newman$-$Janis algorithm \cite{Newman:1965tw} has been applied  to generate rotating solutions in modified gravity theories \cite{Johannsen:2011dh,Bambi:2013ufa,Ghosh:2014pba,Moffat:2014aja,Kumar:2017qws,Kumar:2020hgm,Kumar:2020owy}. The  algorithm was  modified by Azreg-A$\ddot{\text{i}}$nou's non-complexification procedure  \cite{Azreg-Ainou:2014pra,Azreg-Ainou:2014aqa} for generating imperfect fluid rotating solutions in the Boyer$-$Lindquist coordinates from their spherically symmetric static solutions and can also generate generic rotating   black hole solutions. When applied to the static and spherically symmetric solution (\ref{metric}), one obtains the rotating   spacetime --  the metrics depend on the mass ($M$) and spin ($a$) as well a NED parameter $(g)$ that measure potential deviation from the Kerr solution \cite{Kerr:1963ud}, which in the Boyer-Lindquist coordinates  reads
\begin{eqnarray}\label{rotbhtr}
ds^2 & = & - \left( 1- \frac{2Mr^2}{\Sigma \sqrt{r^2+g^2}} \right) dt^2 +
\frac{\Sigma}{\Delta }dr^2 + \Sigma d \theta^2\nonumber
\\ & - & \frac{4aMr^2
}{\Sigma \sqrt{r^2+g^2} } \sin^2 \theta dt \; d\phi + \frac{\mathrm{A}}{\Sigma}\sin^2 \theta d\phi^2,
\end{eqnarray}
with
 \begin{eqnarray*}
& & \Sigma=	r^2 + a^2 \cos^2\theta, \;\;\; \Delta=r^2 + a^2 - 2 \frac{M r^2}{\sqrt{r^2+g^2}},\;\;  \nonumber \\ 
& &\mbox{and}\;\;  \mathrm{A}= (r^2+a^2)^2 - a^2 \Delta \sin^2 \theta.
\end{eqnarray*} 
The metric (\ref{rotbhtr}) includes the Kerr black hole solution as the special case when NED is switched off  ($g=0$), and  the Schwarzschild solution for $g=a=0$.  In that case $M=0$, the  metric (\ref{rotbhtr}) actually is nothing more than  the Minkowski spacetime expressed in spheroidal coordinates.  When only $a=0$,  the rotating   metrics (\ref{rotbhtr}) reduces to  the spherically symmetric   metrics (\ref{metric}) which is a modification of Schwarzschild solution.  It is not difficult to find a range of $M$ and $g$ for which the solution (\ref{rotbhtr}) is a black hole as shown in Fig.~\ref{plot1}. Henceforth, for definiteness, we shall address the solution (\ref{rotbhtr}) as the  magnetically charged rotating black hole.
 \begin{figure*}
 	\begin{tabular}{c c}
 		\includegraphics[scale=0.85]{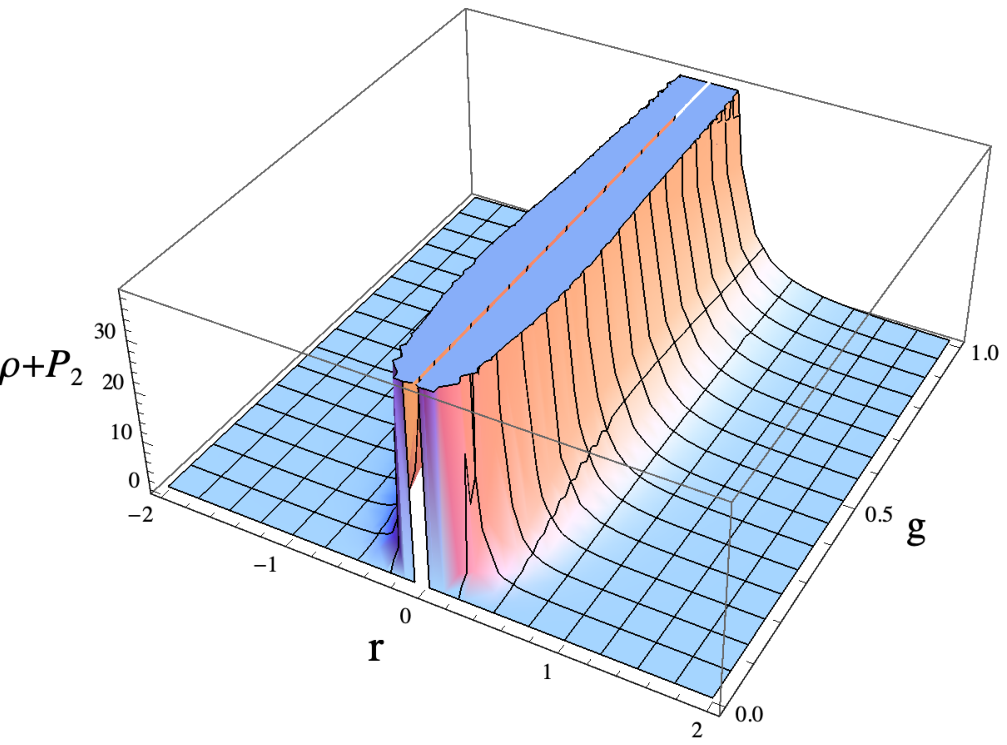}  
 		\includegraphics[scale=0.85]{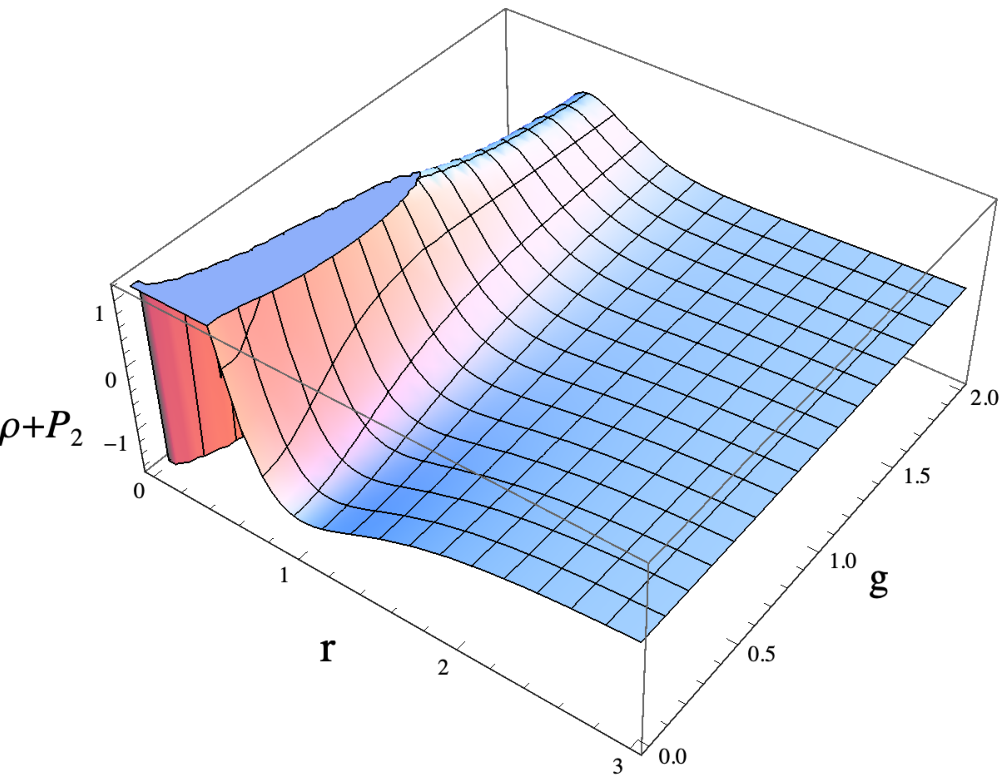}  
 	\end{tabular}
 	\caption{Plots of $\rho+P_2$ vs radius  $r$ and magnetic charge $g$ for $a=0$ (left) and $ a=0.8 $ (right). The reflection  symmetry $r \to -r$  is shown (left top). }; \label{rppp}
 \end{figure*}
In order to further analyze the source associated with the metric (\ref{rotbhtr}), we use orthonormal basis in which energy-momentum tensor is diagonal  \cite{Bambi:2013ufa,Neves:2014aba,Ghosh:2014pba}
\begin{equation}
	e^{(a)}_{\mu}=\left(\begin{array}{cccc}
		\sqrt{\mp(g_{tt}-\Omega g_{t\phi})}& 0 & 0 & 0\\
		0 & \sqrt{\pm g_{rr}} & 0 & 0\\
		0 & 0 & \sqrt{g_{\theta \theta}} & 0\\
		{g_{t\phi}}/{\sqrt{g_{\phi\phi}}} & 0 & 0 & \sqrt{g_{\phi\phi}}
	\end{array}\right),\label{Matrix}
\end{equation}
with $\Omega= g_{t\phi} /{g_{\phi\phi}}$. The components of the energy-momentum tensor in the orthonormal frame reads
\begin{equation}
	T^{(a)(b)} = e^{(a)}_{\mu} e^{(b)}_{\nu} G^{\mu \nu}. \nonumber
\end{equation}
Considering the line element (\ref{rotbhtr}), we can write the components of the respective energy momentum tensor as
\begin{eqnarray}
	& &  \rho = \frac{2r^2m'}{(r^2+a^2)^2} 
	=-P_1, \nonumber \\
	& & P_2 = -\frac{r(r^2+a^2)m''+2a^2 m'}{(r^2+a^2)^2} 
	= P_3,
\end{eqnarray}
where, for brevity, we have used $m(r) = M r/(\sqrt{r^2+g^2})$. To check the weak energy condition, we can choose an appropriate orthonormal basis \cite{Bambi:2013ufa,Neves:2014aba,Ghosh:2014pba} in which the energy momentum tensor reads
\begin{equation}
	T^{(a)(b)} = \mbox{diag}(\rho, P_1,P_2,P_3).
\end{equation}
These stresses vanish for $M=0$, fall off rapidly at large $r$ for $M \neq 0$ .  The weak energy condition requires $\rho\geq0$ and $\rho+P_i\geq0$ ($i=1,\;2,\;3$) \cite{he}. Clearly $\rho>0$ and 
the behaviour of 
$ \rho+P_2 = \rho+P_3$ is depicted in Fig.~\ref{rppp}, which shows the  weak energy conditions for a   black hole are respected  for the spherical case ($a=0$), but may not be prevented when $a\neq0$ (cf. Fig. \ref{rppp}).   The weak energy condition is not 
satisfied for   rotating black holes, but the violation can be very small, depending on the  value of $g$, as shown in the Fig. \ref{rppp}.   However, this happens for mostly all of the the rotating black holes with NED as source \cite{Bambi:2013ufa,Ghosh:2014pba,Kumar:2020hgm}. Despite small violation, such solutions are important from phenomenology and also they are important as astrophysical black holes are rotating.
\begin{figure*}[t]
	\begin{center}	
		\begin{tabular}{c c c c} 
			\includegraphics[scale=0.75]{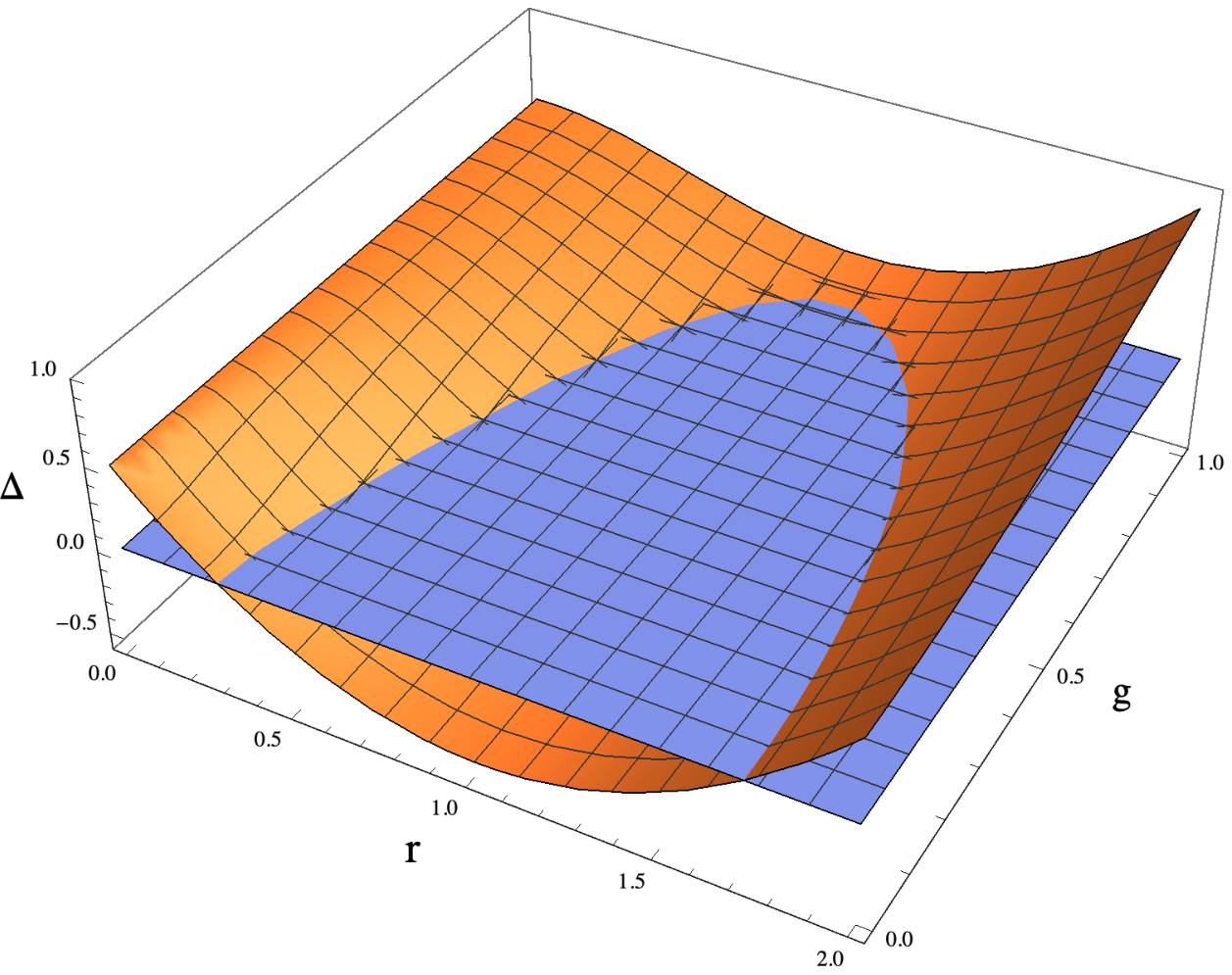}&
			\includegraphics[scale=0.75]{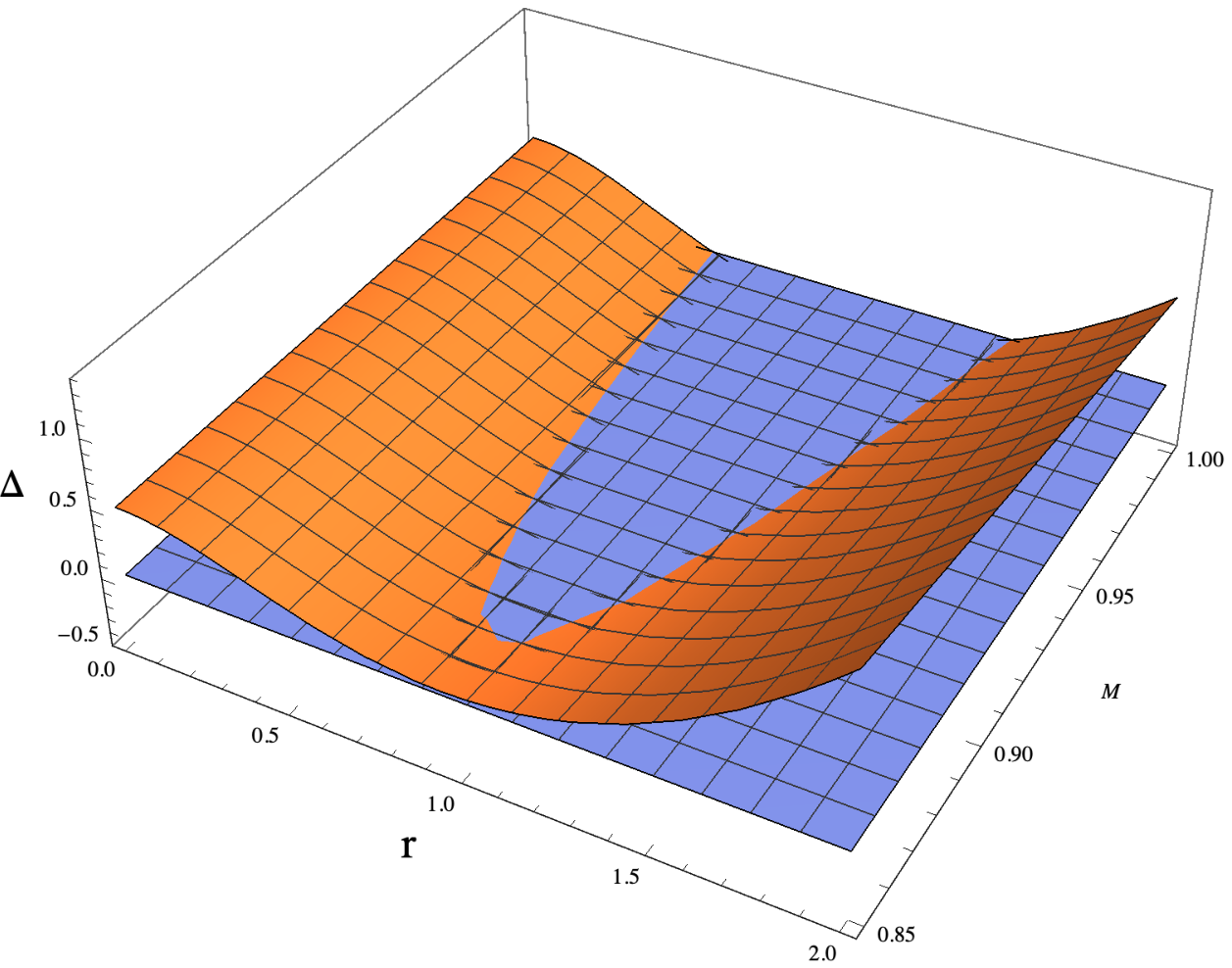}
		\end{tabular}
		\caption{Plot showing the variation of $\Delta$ with $r$ for different values of parameters $g$ and $M$ for black hole for $a=0.7$ (left) and $ a=0.7, g=0.6 $ (right). The blue surface correspond to $\Delta=0$ and values at the intersection points between two surfaces are the radii of horizons. The critical mass for black hole existence is $M_E  \approx 0.866789$ with extremal horizon radius ($r^{H}_-=r^{H}_+ \approx 0.94208$)  for $ a=0.7, g=0.6 $.  For $a=0.7$, we have $g_E \approx 0.9039$ corresponding extremal black hole with  ($r^{H}_-=r^{H}_+ \approx 1.08765$).}\label{plot2}	
	\end{center}   
\end{figure*}

The magnetically charged rotating   black hole metric ~(\ref{rotbhtr}) is independent of $\phi$, and $t$ coordinates and hence admit two Killing vectors, respectively, $\eta_{(t)}^{\mu}=\delta^{\mu}_t$ and $\eta_{(\phi)}^{\mu}=\delta^{\mu}_{\phi}$. Thus, by definition of Killing vectors, test particle four-momentum components associated with translation along $t$ and $\phi$ coordinates are constant of motion. The solution (\ref{rotbhtr}), like the Kerr black hole, is singular at $ \Sigma = 0 $ and at $\Delta=0$. The solution of $\Sigma=0$ is a ring shape curvature singularity. The $\Delta=0$ is a coordinate singularity which determines the horizon -- a null hypersurface of constant $r$, i.e., 
\begin{equation}\label{horizon}
	g^{\mu\nu}\partial _{\mu}r\partial_{\nu}r=0\;\;\; or\;\; g^{rr}=\Delta=0, 
\end{equation} 
where $\partial _{\mu}r$ is the normal to the said hypersurface. The event horizon is a null stationary surface representing the locus of outgoing future-directed null geodesic rays that never manage to reach arbitrarily large distances from the black hole \cite{Hawking:1971vc,he,Poisson:2009pwt}. 

\begin{figure*}[htb]
	\begin{center}	
		\begin{tabular}{c c}
			\includegraphics[scale=0.75]{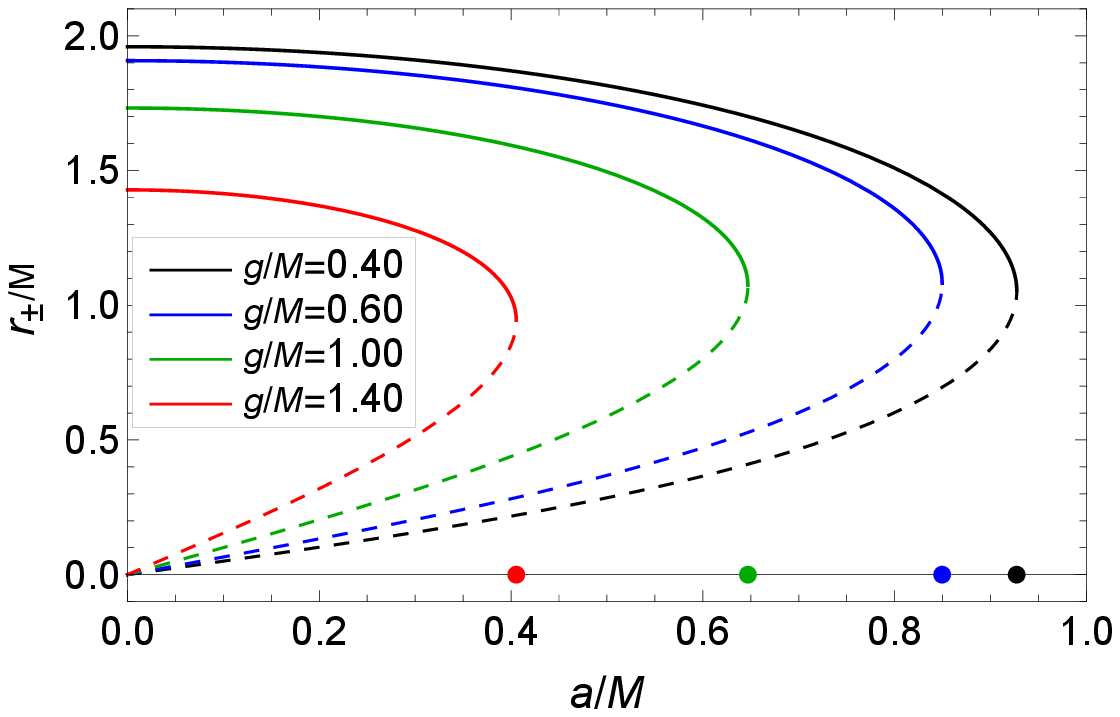}&
			\includegraphics[scale=0.75]{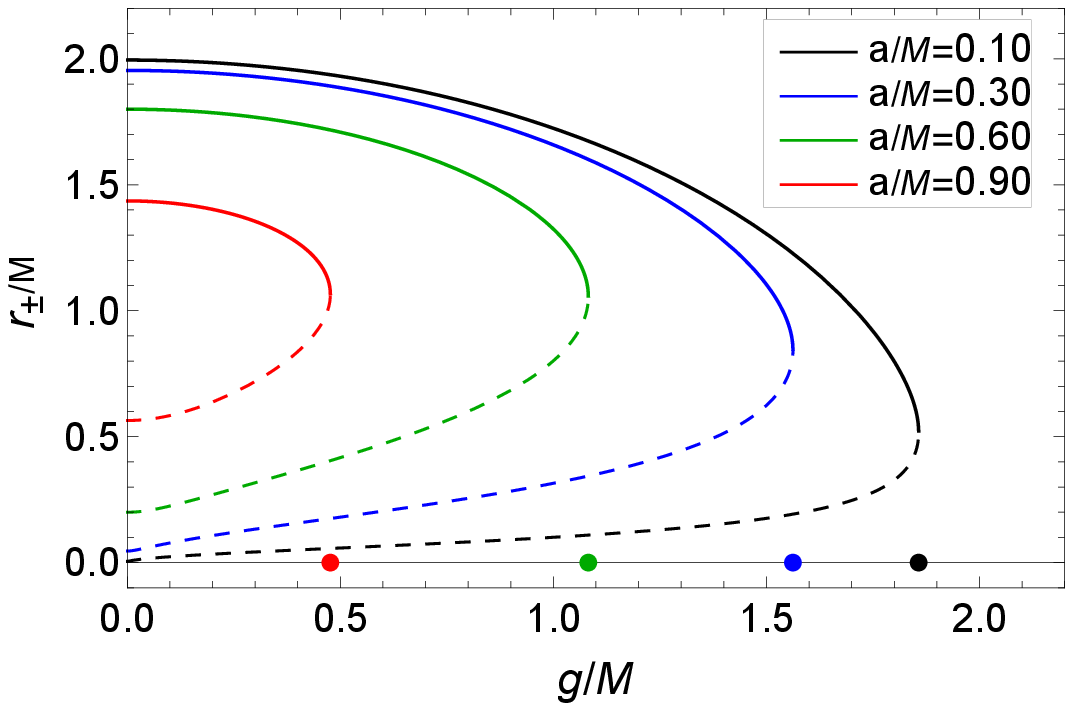}\\
		\end{tabular}
		\caption{Plot showing the variation of event horizon $r_{+}$ (solid lines) and Cauchy horizon $r_{-}$ (dashed lines)  with parameters  $a$ and $g$. Points on the horizontal axis correspond to the critical values for extremal black holes. }\label{plot3}	
	\end{center}   
\end{figure*}

There exist non-vanishing values of parameters $a$ and $g$ for which $\Delta$ has a minimum, and it admits two positive roots $r_{\pm}$ (with $r_{+} \geq r_{-}$).
The positive real roots of the metric component $\Delta=0 $ correspond to the horizons of the black hole relative to the mass distribution. A numerical analysis of the zeros of $\Delta=0$ reveals a critical mass $M_E$ and radius $r_E$ such that  $\Delta=0$  has no zeros if $M<M_E$  -- a geometry of naked singularity,  one double zero at $r = r_E$ if $M=M_E$, i.e.,  there exists an extremal black hole configuration with horizon radius $r_E$, and two simple zeros at $r=r_{\pm}$ if $M<M_E$  (cf. Fig.~\ref{plot2} (right)) -- a  non-extreme black hole with inner and outer horizons $r_{-}$ and $r_{+}$, respectively . 

Similarly, for a given $a$ and $M$, there exists a critical value of $g$, $g_E$, such that $\Delta=0$ has two equal root corresponding to an extremal black hole with degenerate horizons ($r_{-}=r_+=r_E$). When $g<g_E$, $\Delta=0$ has two simple zeros, and has no zeros for $g>g_E$ (cf. Fig.~\ref{plot2} (left)). They, respectively, correspond to a   non-extremal black hole with a Cauchy horizon and an event horizon, and a   no black hole spacetime.  The coordinate  $r$ may take positive as well as negative values. The magnetically charged rotating metric (\ref{rotbhtr}) is symmetric under the reflection $r \to -r$, and the spacetime it describes is composed of two identical portions glued at $ r = 0 $ (cf. Fig.~\ref{rppp} (left)).
\begin{figure*}[htb]
	\begin{center}	
		\begin{tabular}{c c c c}
			\includegraphics[scale=0.75]{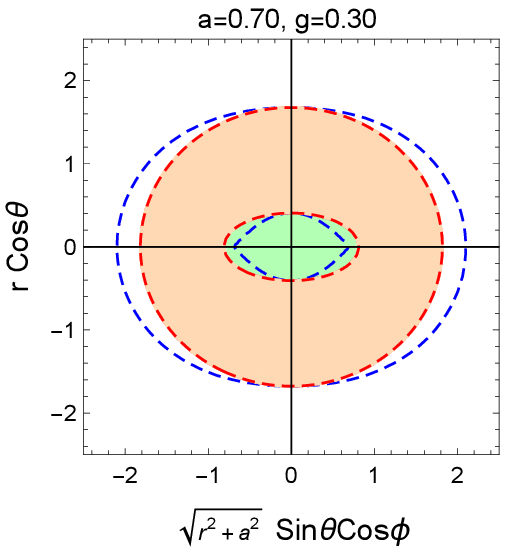}&
			\includegraphics[scale=0.75]{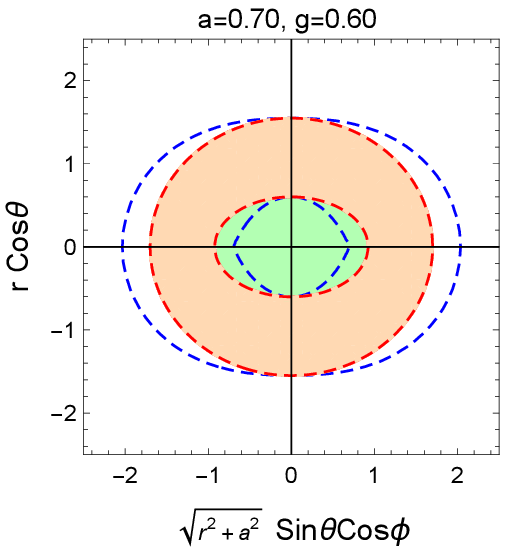}&
			\includegraphics[scale=0.75]{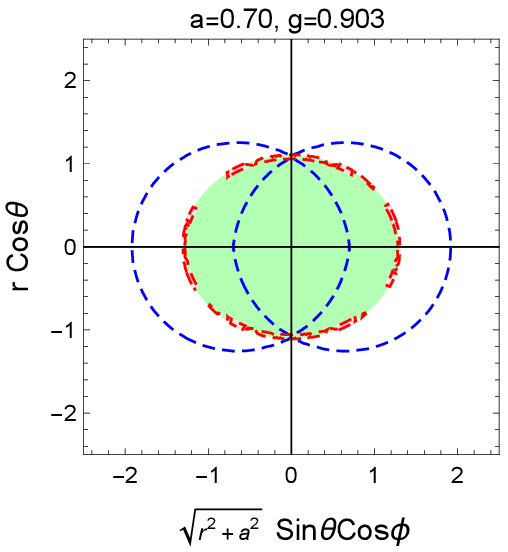}&
			\includegraphics[scale=0.75]{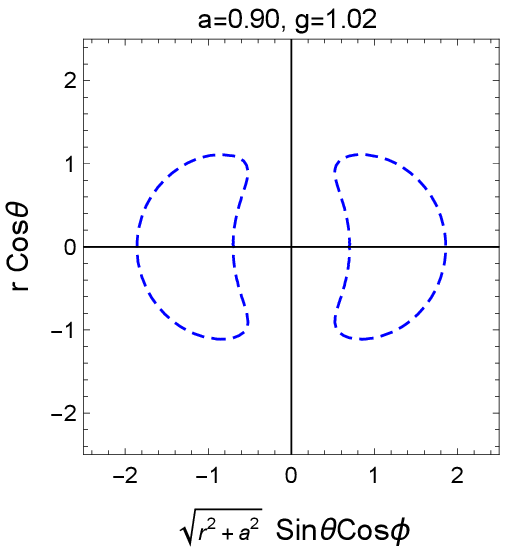}\\
			\includegraphics[scale=0.75]{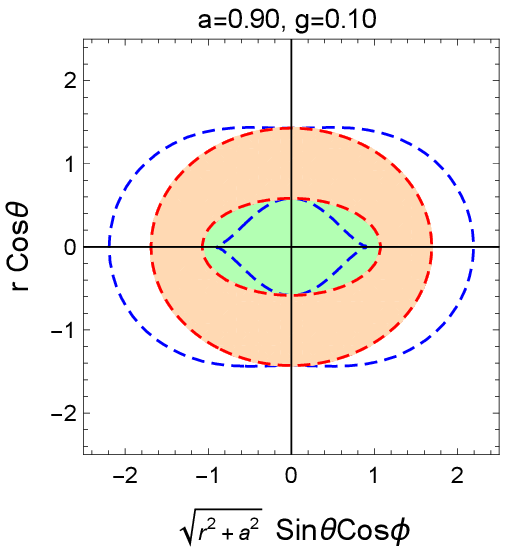}&
			\includegraphics[scale=0.75]{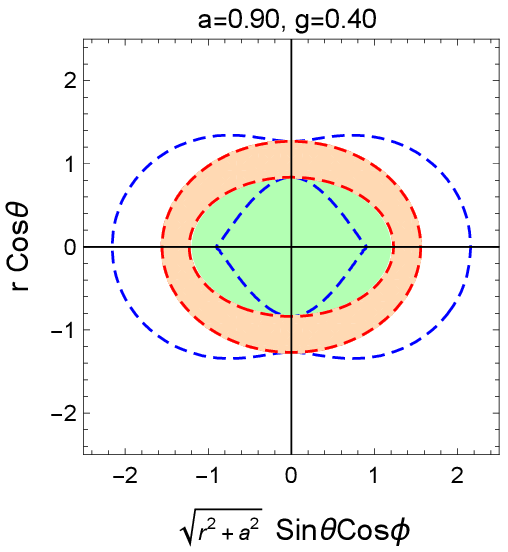}&
			\includegraphics[scale=0.75]{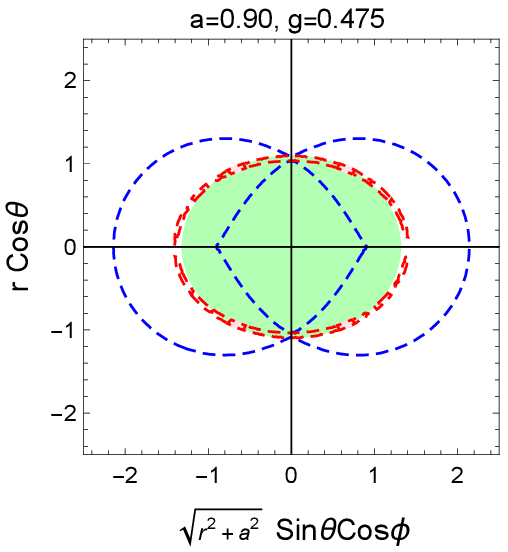}&
			\includegraphics[scale=0.75]{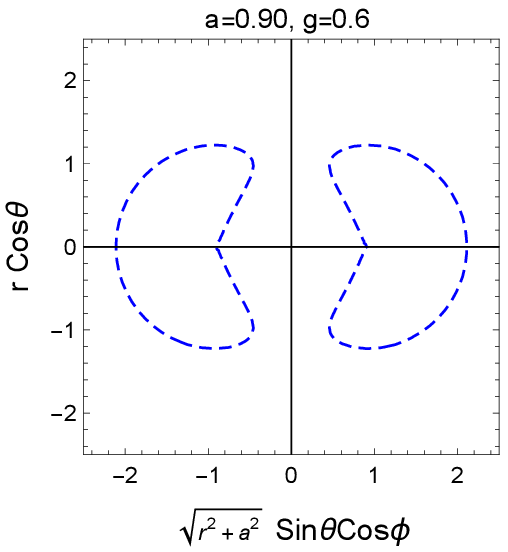}\\
		\end{tabular}
		\caption{The cross-section of the stationary limit surface, the event horizon and the ergosphere variation with parameters  $a$ and  $g$ of the rotating   black holes. The increase in value of the  parameter $g$, for a given $a$, leads to a disconnected event horizon}\label{plot4}	
	\end{center}   
\end{figure*}
The timelike killing vector $\xi^a =(\frac{\partial}{\partial t})^a$
of the solution has norm
\begin{equation}
	\eta_{(t)}^{\mu} \eta_{(t)\mu}^{}=g_{tt}=-\left(\frac{\Delta-a^2\sin^2\theta}{\Sigma}\right),
\end{equation}  
which is null at the static limit surface (SLS), whose locations are, for different $g$,  depicted in Fig.~\ref{plot3}. The region between $r_+^{H}\, < r\, < r_+^{SLS}$ is called \textit{ergosphere}, where the asymptotic time translation Killing field $\xi^a=(\frac{\partial}{\partial t})^a$ becomes spacelike and an observer follow an orbit of $\chi^{\mu}$. The shape of the ergosphere, therefore, depends on the spin $a$, and parameter $g$ (cf. Fig.~\ref{plot4}). Penrose \cite{pc} suggested that energy can be extracted from a black hole that relies on the presence of an \textit{ergosphere}.  Thus, the dependence of ergosphere on the NED parameter $g$, in turn, is likely to have an impact on energy extraction being investigated separately.  The vacuum state is obtained by letting horizons size go to zero or by making black hole disappear this amounts to $r \rightarrow \infty$.  One thus conclude that the solution is asymptotically flat as the metric components approaches those  of the Minkowski  spacetime in spheroidal coordinates.
\begin{figure}
	\includegraphics[scale=0.77]{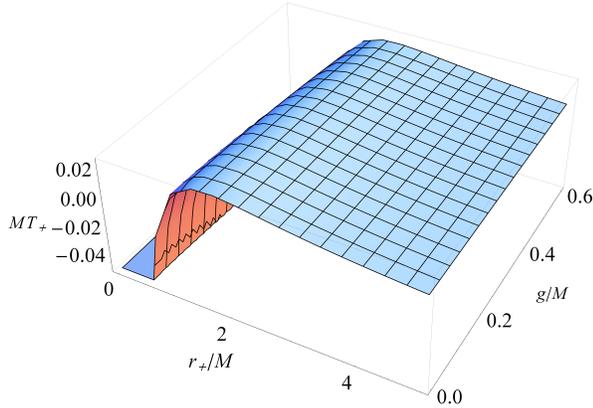}	
	\caption{ Plot of temperature $T_+$  vs horizon radius $r_+$.} \label{ptemp}	
\end{figure}

\section{Thermodynamics and  Komar conserved quantities}\label{sec3}
Next, the zero angular momentum observers (ZAMO)  are the stationary observers relative to spatial infinity, but due to frame dragging gain non-zero angular velocity $\omega = {d\phi}/{dt}$. The $\omega$ vanishes for ZAMO at infinity, but in the case, it is non-zero and position-dependent \cite{Poisson:2009pwt}. For the magnetically charged rotating   black hole a stationary observer outside the event horizon, moving with zero angular momentum with respect to an observer at spatial infinity can rotate with the black hole with an angular velocity given by 
\begin{equation}\label{zamo}
	\omega=-\frac{g_{t\phi}}{g_{\phi\phi}}={\frac {a \left( r^2+a^2-\Delta\right) }{ {a}^{4} \sin^{4} \theta -{a}^{2} \left( \Delta-2\,\Sigma \right) \sin^{2} \theta+{\Sigma}^{2}}}
\end{equation}
The limit of the ZAMO angular velocity (\ref{zamo}) at the horizon $	\omega_+=a/(r_+^2+a^2)$, i.e.,  every point of the horizon has the same angular velocity (as measured at infinity). The Killing vectors $\eta^{\mu}_{(t)}$ or $\eta^{\mu}_{(\phi)}$ are not the  generators of the stationary black hole horizon, instead, it is their linear combination \cite{Chandrasekhar:1992} as
\begin{equation}
	\chi^{\mu}=\eta^{\mu}_{(t)}+\Omega \eta^{\mu}_{(\phi)},
\end{equation}
such that $\chi^{\mu}$ is globally time-like outside the event horizon, though it is Killing vector only at the horizon \cite{Chandrasekhar:1992}. Using the null property of Killing vector $\chi^{\mu}=\eta_{t}^{\mu}+\Omega\eta_{\phi}^{\mu}$, we can calculate the rotational velocity  
\begin{equation}
	\Omega = \frac{\left[\pm\Sigma\sqrt{\Delta}+a\sin\theta(r^2+a^2-\Delta)\right]}{\left[a^4\sin^4\theta -a^2(-2\Sigma+\Delta)\sin^2\theta+\Sigma^2\right]\sin\theta};
\end{equation}
at horizon $\Delta=0$, we get the horizon rotational frequency reads as
\begin{equation}\label{frequency}
	\Omega_+=\frac{a}{r_+^2+a^2}.
\end{equation}
and corresponds to the Kerr black hole value \cite{Poisson:2009pwt,Chandrasekhar:1992} and at the horizon $	\omega_+=\Omega_+$. $\chi^{\mu}$ is not a Killing vector for arbitrary $r$ because the angular velocity ($\Omega$), even for a fixed $r$, depends on the polar angle $\theta$. 
\begin{figure*}
	\begin{tabular}{c c}
		\includegraphics[scale=0.95]{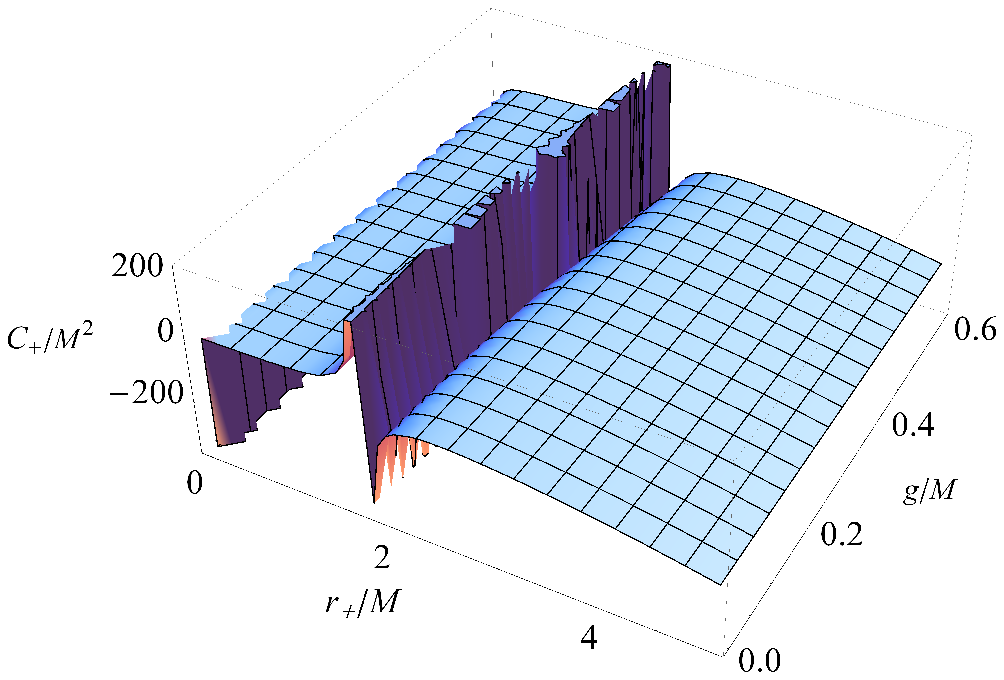}  \hspace*{0.2cm}&
		\includegraphics[scale=0.90]{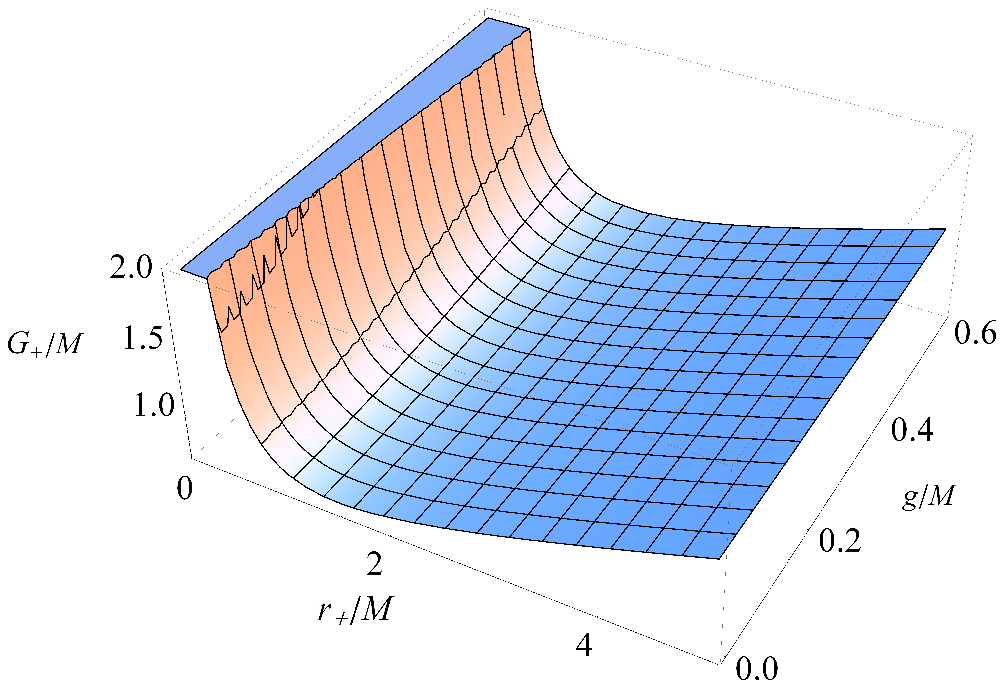}  
	\end{tabular}
	\caption{Plots of $C_{+}$ (left) and  $G_{+}$ (right) vs radius  $r_+$ and magnetic charge $g$ for $a=0.8$. }\label{stab}
\end{figure*}
The black hole mass and angular momentum correspond to the conserved quantities associated with the $\eta^{\mu}_{(t)} $ and $\eta^{\mu}_{(\phi)}$, respectively. Let us consider a space-like hypersurface $\Sigma_t$, extending from the event horizon to spatial infinity, which is a surface of constant $t$ with unit-normal vector $n_{\mu}$  \cite{Chandrasekhar:1992,Wald}.  The effective mass reads \cite{Komar:1958wp}
\begin{equation}
	M_{\text{eff}}=-\frac{1}{8\pi}\int_{S_t}\nabla^{\mu}\eta^{\nu}_{(t)}dS_{\mu\nu},\label{mass}
\end{equation}
where $S_t$ is the two-boundary of the hypersurface $\Sigma_t$ and is a constant-$t$ and constant-$r$ surface with unit outward normal vector $\sigma_{\mu}$, $dS_{\mu\nu}=-2n_{[\mu}\sigma_{\nu]}\sqrt{h}d^2\theta$ is the surface element of $S_t$, $h$ is the determinant of ($2\times 2$) metric on $S_t$ and 
\begin{equation}
	n_{\mu}=-\frac{\delta^{t}_{\mu}}{|g^{tt}|^{1/2}},\qquad \sigma_{\mu}=\frac{\delta^{r}_{\mu}}{|g^{rr}|^{1/2}},
\end{equation}
are, respectively, timelike and spacelike unit outward normal vectors. Thus mass integral Eq.~(\ref{mass}) turned into integral over closed 2-surface at infinity
\begin{align}
	M_{\text{eff}}
	=& \frac{1}{4\pi}\int_{0}^{2\pi}\int_{0}^{\pi}\frac{\sqrt{g_{\theta\theta}g_{\phi\phi}}}{|g^{tt}g^{rr}|^{1/2}}\left(g^{tt}\Gamma^{r}_{tt}+g^{t\phi}\Gamma^{r}_{t\phi} \right)d\theta d\phi.
\end{align}
Using the metric (\ref{rotbhtr}), the effective mass of the magnetically charged rotating black holes read
\begin{align}
	M_{\text{eff}}=&\frac {M}{a \left( {g}^{2}+{r}^{2} \right)^{3/2} } \Big[ra \left( {g}^{2}+{r}^{2} \right) \nonumber\\
	& - {g}^{2} \left( {a}^{2}+{r}^{2} \right) \arctan
		\left( {\frac {a}{r}} \right)
		\Big], \label{mass1}
\end{align}
which is corrected due to the NED, and goes over to the Kerr black hole ($g=0$) value that is $M_{\text{eff}}=M$. Next, we utilize the spacelike Killing vector $\eta^{\mu}_{(\phi)}$ to calculate the effective angular momentum  \cite{Komar:1958wp}
\begin{equation}
	J_{\text{eff}}=\frac{1}{16\pi}\int_{S_t}\nabla^{\mu}\eta^{\nu}_{(\phi)}dS_{\mu\nu},\label{ang}
\end{equation}
using the definitions of surface element, Eq.~(\ref{ang}) recast as
\begin{align}
	J_{\text{eff}}=&-\frac{1}{8\pi}\int_{0}^{2\pi}\int_{0}^{\pi}\nabla^{\mu}\eta^{\nu}_{(t)}n_{\mu}\sigma_{\nu}\sqrt{h}d\theta d\phi,\nonumber\\
	=& \frac{1}{8\pi}\int_{0}^{2\pi}\int_{0}^{\pi}\frac{\sqrt{g_{\theta\theta}g_{\phi\phi}}}{|g^{tt}g^{rr}|^{1/2}}\left(g^{tt}\Gamma^{r}_{t\phi}+g^{t\phi}\Gamma^{r}_{\phi\phi} \right)d\theta d\phi.
\end{align}
 Upon using the metric (\ref{rotbhtr}) and  integrating,  the effective angular momentum for the magnetically charged rotating black holes becomes 
\begin{align}
	J_{\text{eff}}=&\frac {M}{2\,{a}^{2}  \left(\ {g}^{2}+{r}^{2}
			\right)^{3/2}} \Big[ \left( 2\,{a}^{3}+a{g}^{2
		} \right) {r}^{3}+3\,{a}^{3}{g}^{2}r  \nonumber \\
	 &-{g}^{2} \left( {a}^{2}+{r}^{2} \right) 
		^{2}\arctan \left( {\frac {a}{r}} \right) \Big] 
	\label{ang1}
\end{align}
In the neutral limit, $g \to 0$, the effective angular momentum Eq.~(\ref{ang1}) restore the Kerr black hole value $J_{\text{eff}}=Ma$.

The Komar conserved quantity at the event horizon, associated with $\chi^{\mu}= \eta^{\mu}_{(t)}+\Omega \eta^{\mu}_{(\phi)}$, reads as \cite{Komar:1958wp}
\begin{eqnarray}
\mathcal{K}_{\chi}&=&-\frac{1}{8\pi}\int_{S_t}\nabla^{\mu}\chi^{\nu}dS_{\mu\nu}, \nonumber\\
	&=&-\frac{1}{8\pi}\int_{S_t}\nabla^{\mu}\left( \eta^{\mu}_{(t)}+\Omega \eta^{\mu}_{(\phi)}\right)dS_{\mu\nu}.
\end{eqnarray}
Using Eqs.~(\ref{mass1}) and (\ref{ang1}), we obtain
\begin{eqnarray}
\mathcal{K}_{\chi}&=&M_{\text{eff}}-2\Omega J_{\text{eff}},\nonumber\\
	&=&\frac {Mr_+ \left( r_+^{4} -2\,{a}^{2}{g}^{2}-{a}^{2}r_+^{2}\right) }{\left({a}^{2}+r_+^{2} \right) \left( {g}^{2}+r_+^{2} \right)^{3/2} }.\label{ST}\end{eqnarray}
Next, we analyse the thermodynamical quantities associated with magnetically charged rotating black hole  metric (\ref{rotbhtr}). An extended form of zeroth law of black hole mechanics implies that the surface gravity $\kappa$, the angular velocity $\Omega$ and the electrostatic potential are all locally defined on the horizon and they are always constant over the horizon of any stationary black hole \cite{Kumar:2017qws}. Area of the black hole horizon can be calculated by the metric components \cite{Poisson:2009pwt}
\begin{equation}
	A_H=\int_{0}^{2\pi}d\phi\int_{0}^{\pi}\sqrt{g_{\theta\theta}g_{\phi\phi}}d\theta,
\end{equation}
which upon integration leads to 
$ A_H=4\pi(r_+^2+a^2)$, $r_+$ is outer horizon radius. The universal area law of black hole mechanics gives the entropy of black hole as 
\begin{equation}\label{entropy}
S_{+}=\frac{A_H}{4}=\pi (r_+^2+a^2).
\end{equation}
The black holes are characterized by their mass $(M_+)$, which  can be expressed in terms of horizon radius ($r_+$) and reads as
\begin{equation}\label{BHMass}
M_{+} = {\frac {{a}^{2}+r_+^{2}}{2\,r_+^{2}}\sqrt {{g}^{2}+r_+^{2}}}.
\end{equation}
In the limiting case of $g=0$ the  above expression reduces to $M_{+}= ({{a}^{2}+r_{+}^{2}})/{(2\,r_+)}$, corresponding to the Kerr black hole \cite{Poisson:2009pwt}.
Since, black hole behaves as a thermodynamical entity whose temperature $T$ can be calculated from surface gravity $\kappa$ evaluated at Killing horizon through $\kappa^2=-\frac{1}{2}\chi^{\mu;\nu}\chi_{\mu;\nu}.$
Hawking showed that the black hole temperature is determined by
\begin{eqnarray}
	T_+= && 
	\frac{\kappa}{2\pi}=\frac{\Delta'(r_+)}{4\pi(r_+^2+a^2)} \nonumber \\
&& 	= {\frac {r_+^{4} -2\,{a}^{2}{g}^{2}-{a}^{2}r_+^{2}}{4\pi r_+ \left( {g}^{2}	+r_+^{2} \right) \left( {a}^{2}+r_+^{2} \right) }}, 
	\label{temp}
\end{eqnarray}
which in the absence of the NED ($ g=0 $), yields
\begin{eqnarray}
	T_+^{K}&=& {\frac {r_{+}^{2} - {a}^{2}}{4 \pi\,r_+ \left( {a}^{2}+r_{+}^{2} \right) }}, 
	\label{tempK}
\end{eqnarray}
the temperature of the Kerr black hole \cite{Kumar:2017qws,Kumar:2020hgm}. It is evident from Fig.~\ref{ptemp}, the Hawking temperature of the magnetically charged rotating black holes grows to a maximum $T^{max}_+$ then drops to zero temperature at $r=r_E$ and becomes negative at a small radius.  A local maximum of the  Hawking temperature occurs at the critical radius $r_c$; the maximum value of the Hawking temperature decreases with an increase in the values of the  NED parameter $g$. The Hawking temperature of the magnetically charged rotating black hole spacetime increases with the horizon size, i.e., with the mass (cf. Fig.~\ref{ptemp} ) when horizon radii in the range $r_E \leq r \leq r_C$. It implies a positive heat capacity and thermodynamical stability of the black hole. At $r_C$,  a divergence of the heat capacity indicates a black hole thermodynamical phase transition between the phase of positive and that of negative heat capacity.

From the classical electrodynamics we can calculate the potential $ \Phi$ associated with black hole charge $g$ \cite{Chen:2008ra, Sekiwa:2006qj}. Therefore, the differential form of the first law of black hole thermodynamics can be written as
\begin{equation}\label{flaw}	
dM=TdS+\Omega dJ+\Phi dg
\end{equation}
Furthermore, using this we can calculate the extensive quantity associated with black hole, i.e. temperature, angular velocity and electrostatic potential through
\begin{equation}\label{first}
	T=\left(\frac{dM}{dS}\right)_{(J,g)},\;
\Omega=\left(\frac{dM}{dJ}\right)_{(S,g)},\;
\Phi=\left(\frac{dM}{dg}\right)_{(S,J)}.
\end{equation}
Lastly, Eqs (\ref{ST}), (\ref{temp}) and (\ref{entropy}),  leads to 
\begin{equation}
	\mathcal{K}_{\chi}=M_{\text{eff}}-2\Omega J_{\text{eff}}=2S_+T_+.
\end{equation}
Therefore, the Komar conserved quantity corresponding to the null Killing vector at the event horizon $\chi^{\mu}$ is twice the product of the black hole entropy and the horizon temperature and hence satisfy the Smarr formula \cite{Smarr:1972kt,Bardeen:1973gs,Kumar:2017qws}.

Finally we analyse the thermodynamic stability of the magnetically charged rotating black holes which requires the  study  of its heat capacity, which is  defined as \cite{Cai:2003kt,Sahabandu:2005ma}
\begin{equation}\label{sh_formula}
	C_+ = \frac{\partial{M_+}}{\partial{T_+}}= \left(\frac{\partial{M_+}}{\partial{r_+}}\right)
	\left(\frac{\partial{T_+}}{\partial{r_+}}\right)^{-1}. 
\end{equation} 
The global stability of the black hole can be deduced from the behaviour of its free energy. The Gibb's free energy of black hole in the canonical ensemble is obtained as \cite{Altamirano:2014tva,Carlip:2003ne}
\begin{eqnarray}\label{fe}
	G_+&=&M_+-T_+S_+.
\end{eqnarray} 
On using  Eqs.~(\ref{entropy}), (\ref{BHMass}) and (\ref{temp}) into Eqs.~(\ref{sh_formula}) and (\ref{fe}), we get the expression, respectively, for heat capacity and Gibb's free energy. The graphic results, depicted in Fig. \ref{stab},  might be more enlightening than the analytical expressions as they are complicated and lengthy to present here. However, as a consistency check, it is easily observed that  the expression for heat capacity and Gibb's free energy for the  Kerr black hole  recovered when $g=0$, and takes a straightforward form  \cite{Kumar:2017qws,Kumar:2020hgm,Altamirano:2014tva}
\begin{eqnarray*}
	C_{+}^K & =	&{\frac {2\pi\, \left( r_+^{2} -{a}^{2}\right)  \left( {a}^{2}+r_+^{2}	\right) ^{2}}{{a}^{4}+4\,{a}^{2}{r_+}^{2}-r_+^{4}}}, \nonumber \\
	G_{+}^K & = &	\frac { \left( 3{a}^{2}+r_+^{2} \right)}{4\,r_{+}}.
\end{eqnarray*}
The positivity of heat capacity $C_+>0$ of the black hole is sufficient to state that the black hole is thermodynamically stable to local thermal fluctuation. A black hole, at some stage, due to thermal fluctuations, absorbs more radiation than it emits, which leads to positive heat capacity \cite{Cai:2003kt,Sahabandu:2005ma, Ghosh:2014pga}.  When the specific heat is positive, an increase in the black hole temperature will increase the entropy, thereby giving a stable thermodynamic configuration.  Fig. \ref{stab} shows that heat capacity, for a given value of $g$ and $a$,  is discontinuous at a critical radius $r^{C}_+$. Further, we noticed that the heat capacity flips its sign around $r^{C}_+$. Thus, we can say the black hole is thermodynamically stable for  $r_1<r_+ < r^{C}_+$ where ${C}_+>0$, whereas it is thermodynamically unstable for $r_+>r_+^C$ region wherein ${C}_+<0$, and there is a second-order phase transition at $r_+=r^{C}_+$. Thus, we can say that the heat capacity is negative for a larger black hole with $r_+>r_+^C$, positive for the black hole in the region $r_1<r+<r_+^C$  and again negative for minimal radius $r<r_1$. It means that the smaller size black holes are thermodynamically stable locally \cite{Cai:2003kt,Sahabandu:2005ma,Ghosh:2014pga}. 

One can analyze the free energy to discuss the global thermodynamical stability of black hole \cite{Altamirano:2014tva}. If we consider that the black hole is in thermodynamical equilibrium with a reservoir such that it exchanges only mass, then in the preferred phase, the free energy will be minimum. As we all know, the thermodynamic state with lower Gibbs free energy is more stable. We depict the Gibbs free energy for various values of parameter $g$ and $a$  in Fig.~\ref{stab}. We note that the free energy is positive for the entire parameter space $(r_+,\;g)$. Interestingly, the global minima of the Gibbs free energy occurs for $r<r_c$, as shown in Fig. \ref{stab} where specific heat is also positive. Consequently, in a fixed charge, canonical ensemble, smaller magnetically charged rotating black holes are thermodynamically preferred than the large black holes \cite{Altamirano:2014tva}. The larger black holes ($r_+>r_c$) have more positive Gibbs free energy and negative specific heat and, hence, thermodynamically unstable. The Hawking-Page-type phase transition is not possible as the free energy  $G_{+}>0$ for all $r_+$ as depicted in Fig.~\ref{stab}. 

\section{Radiating   rotating metric}
Next, we add radiation  by rewriting the static solution
(\ref{rotbhtr}) in  terms of the Eddington-Finkelstein coordinates
$(v,r,\theta,\phi)$ \cite{Chandrasekhar}:
\begin{equation}
	v = t+\int \frac{r^2+a^2}{\Delta}, \;
	\bar{\phi} = \phi + \int \frac{a}{\Delta},
\end{equation}
and allow mass $M$ and parameter $g$ to be function of advanced time $v$, and hereafter dropping bar, we get
\begin{eqnarray}\label{rotbhvr}
	ds^2 & = & -\left( 1- \frac{2M(v)r^2 }{\Sigma \sqrt{r^2+g(v)^2}} \right) dv^2
	+ 2 dv dr + \Sigma d \theta^2  \nonumber \\ & - & \frac{4aM(v)r^2
	}{{\Sigma \sqrt{r^2+g(v)^2}}} \sin^2 \theta dv  d\phi -2 a \sin^2 \theta dr
	d\phi  \nonumber \\ & + &  \mathrm{A'}\sin^2 \theta d\phi^2.
\end{eqnarray}
\begin{eqnarray*}
	& & \Delta'=r^2 + a^2 - 2 \frac{M (v) r^2}{\sqrt{r^2+g(v)^2}},\;\;  \nonumber \\ 
	& &\mbox{and}\;\;  \mathrm{A'}= (r^2+a^2)^2 - a^2 \Delta' \sin^2 \theta. 
\end{eqnarray*} 
The stress-tensor components associated with rotating radiating black holes (\ref{rotbhvr}) have the same mathematical form as that of the analogous stationary solution (\ref{rotbhtr}) but with mass $M(v)$. However, it (\ref{rotbhvr}) also has additional stresses corresponding to the energy-momentum tensor of null radiation \cite{Carmeli}. The solution (\ref{rotbhvr}) describe an exterior of radiating objects, regaining Carmeli solution (or rotating Vaidya solutions) \cite{Carmeli}  for $g=0$, and also in the limit $g=a=0$ Vaidya solution \cite{pc}. Thus, the radiating rotating solution (\ref{rotbhvr}) is a natural generalization of the stationary rotating solution (\ref{rotbhtr}), but it is Petrov type-II with a twisting, shear free, null congruence the same as for stationary rotating solution, which is of Petrov type $ D $. Thus, the rotating radiating black hole (\ref{rotbhvr}) bears the same relation to a stationary black hole (\ref{rotbhtr}) as does the Vaidya solution to the Schwarzschild solution.  

\section{Conclusions}
This paper obtains an exact static spherically symmetric black hole spacetimes by solving Einstein equations coupled with NED, choosing suitable Lagrangian density, which encompasses the Schwarzschild as a special case the absence of NED ($ g = 0 $). In turn, we also obtained a rotating counterpart using the modified Newman and Janis algorithm.  The source, if it exists, is the same for both a black hole in GR and its rotating counterpart, e.g., the vacuum for both Schwarzschild and Kerr black holes, and charge for Reissner-Nordstrom and Kerr-Newman black holes. However, the NED charged  rotating counterpart (\ref{rotbhtr})  matter have some additional stresses than spherical solution (\ref{metric}).

The magnetically charged rotating black hole metric (\ref{rotbhtr})  is asymptotically flat and encompasses Kerr ($g=0$) and Schwarzschild ($g=0,a=0$) black holes, and like the Kerr black hole, our solutions can represent black holes with inner and outer horizons, an extreme black hole or naked singularity depending on the choice of the parameters. Interestingly, for a fixed value of $a$ and $g$, we found that there is an extreme value for nonlinear parameter mass $M_E$ and radius $r_E$ such that if $M>M_E$  with two horizons, a black hole with degenerate horizons for  $M=M_E$ and naked singularity for $M<M_E$.  The ergosphere area increases and, thereby, can have interesting consequences on the astrophysical Penrose process. Violations of the weak energy conditions characterize the magnetically charged rotating metric ($a\neq0$), but the same holds for the spherical black holes ($a=0$).  

Further, thermodynamic quantities such as the Hawking temperature, heat capacity, and Gibbs free energy have been derived and plotted. The phase transition is characterized by the divergence of heat capacity at a critical radius $r_c$.  It would be essential to investigate how these black holes with positive heat capacity ($C_+>0$) would emerge from thermal radiation through a phase transition. Despite the complicated magnetically charged rotating metric (\ref{rotbhtr}), using the Komar prescription, we analytically derived the exact expressions for conserved mass $M_{\text{eff}}$ and angular momentum $J_{\text{eff}}$, valid at any radial distance. Furthermore,  the NED  significantly altered these conserved quantities compared with those for the Kerr black hole, which are regained in the limit $g=0$. We have found the null Killing vector $\chi^{\mu}$ at the event horizon, and calculated the corresponding Komar conserved quantity $\mathcal{K}_{\chi}$. Interestingly, $\mathcal{K}_{\chi}$ is found to be twice the product of entropy and temperature of the black hole $\mathcal{K}_{\chi}=2S_+T_+$ and hence satisfies the Smarr formula.

To conclude, a new magnetically charged rotating solution to modified gravity theory opens new opportunities for finding ways to test these theories against astrophysical observations. The analysis of possible mechanisms for extracting rotational energy from the black hole deserves further study. Also, black hole stability needs to be addressed to understand these objects' physical relevance. It would also be interesting to consider the M87* black hole's shadow observational results by EHT to put constraints on the magnetically charged rotating black holes. Further, If we take into account the AdS background for this magnetically charged rotating black hole, there should be interesting phase structure and critical phenomena.  Such investigations now have a clear astrophysical relevance; we hope to be reporting on these issues shortly. 

\begin{acknowledgements} 
We would like to thank the Science and Engineering Research Board, Department of Science and Technology for the ASEAN project IMRC/AISTDF/CRD/2018/000042, and Shaqat Ul Islam for help in plots. R.K.W. would like to thanks UKZN and NRF for the postdoctoral fellowship.
\end{acknowledgements}

\end{document}